\documentclass[useAMS]{mn2e}
 \usepackage{times}
 \usepackage{graphicx}
 \usepackage{epsfig}
 \usepackage{amssymb}
 \usepackage{media9}
 \usepackage{hyperref}

 \title[A transient trojan-horseshoe companion to Venus]
       {Asteroid 2012 XE$_{133}$: a transient companion to Venus}

 \author[C. de la Fuente Marcos and R. de la Fuente Marcos]
        {C.~de~la~Fuente~Marcos\thanks{E-mail: nbplanet@fis.ucm.es}
          and
         R. de la Fuente Marcos \\
         Universidad Complutense de Madrid,
         Ciudad Universitaria, E-28040 Madrid, Spain}
 \date{Accepted 2013 March 10.
       Received 2013 March 8;
       in original form 2013 January 7}
 \pubyear{2013}
 
\begin{document}
  \maketitle

  \begin{abstract}
     Apart from Mercury that has no known co-orbital companions, Venus 
     remains as the inner planet that hosts the smallest number of known 
     co-orbitals (two): (322756) 2001 CK$_{32}$ and 2002 VE$_{68}$. Both 
     objects have absolute magnitudes 18 $<H<$ 21 and were identified as 
     Venus co-orbitals in 2004. Here, we analyse the orbit of the 
     recently discovered asteroid 2012~XE$_{133}$ with $H$= 23.5 mag to 
     conclude that it is a new Venus co-orbital currently following a 
     transitional trajectory between Venus' Lagrangian points L$_5$ and 
     L$_3$. The object could have been a 1:1 librator for several 
     thousand years and it may leave the resonance with Venus within the 
     next few hundred years, after a close encounter with the Earth. Our 
     calculations show that its dynamical status as co-orbital, as well 
     as that of the two previously known Venus co-orbitals, is controlled 
     by the Earth-Moon system with Mercury playing a secondary role. The 
     three temporary co-orbitals exhibit resonant (or near-resonant) 
     behaviour with Mercury, Venus and the Earth and they follow rather 
     chaotic but similar trajectories with e-folding times of the order 
     of 100 yr. Out of the three co-orbitals, 2012 XE$_{133}$ currently 
     follows the most perturbed path. An actual collision with the Earth 
     during the next 10\,000 yr cannot be discarded, an encounter at 
     0.005 au may take place in 2028 but even closer encounters are 
     possible within that time frame. Extrapolation of the number 
     distribution of Venus co-orbitals as a function of the absolute 
     magnitude suggests that dozens of objects similar to 2012 XE$_{133}$ 
     could be transient companions to Venus. Some additional objects that 
     were or will be transient co-orbitals to Venus are also briefly 
     discussed. 
  \end{abstract}


  \section{Introduction}
     Early calculations carried out by Mikkola \& Innanen (1992) showed that the existence of asteroids following co-orbital motion with 
     Venus in the form of tadpole or horseshoe orbits stable for a few Myr was possible. Numerical integrations of the orbits of Near-Earth 
     Objects (NEOs) revealed that some of them can eventually follow co-orbital motion with Venus; for example, (4660) Nereus (Michel 1997) 
     and (99907) 1989 VA (Christou 2000). In his study, Christou also predicted a quasi-steady-state flux of such objects moving in and out 
     of the co-orbital regime. Tabachnik \& Evans (2000) and Brasser \& Lehto (2002) used simulations to show that Venus trojans would be 
     stable in low-inclination orbits. These theoretical findings soon gained observational support with the discovery and subsequent 
     identification of two present-day co-orbital companions to Venus: (322756) 2001 CK$_{32}$, a horseshoe-quasi-satellite orbiter (Brasser 
     et al. 2004), and 2002 VE$_{68}$, a quasi-satellite (Mikkola et al. 2004). Both are relatively small and their respective absolute 
     magnitudes, $H$, are 18.9 and 20.3. The existence of a transient population of NEOs trapped in a 1:1 mean motion resonance with Venus 
     has been explained by Morais \& Morbidelli (2006) within the steady-state scenario envisioned by Bottke et al. (2000, 2002) where NEOs 
     are constantly being supplied from the main asteroid belt. In their paper, it is predicted that the number of Venus co-orbital NEOs 
     with $H <$ 18 and $H <$ 21 is 0.14$\pm$0.03 and 1.6$\pm$0.3, respectively. Taking into account the two objects already known, this 
     translates into completeness for $H <$ 21. If, as Morais \& Morbidelli (2006) suggest in their work, current surveys have reached 
     completeness at such small sizes, no objects in similar orbits and brighter than $H$ = 21 should be observed in the future. On the 
     other hand, numerical simulations indicate that the existence of present-day primordial Venus co-orbitals is very unlikely due to the 
     presence of multiple secular resonances (Scholl, Marzari \& Tricarico 2005).
     \hfil\par
     In this paper, we show that the recently discovered asteroid 2012 XE$_{133}$ is a new transient Venus co-orbital with $H$ = 23.5 which 
     is consistent with the Morais \& Morbidelli (2006) predictions. The object was originally selected as a candidate because of its small 
     relative semi-major axis, $|a - a_{Venus}| \sim$ 0.0004 au, and we use $N$-body calculations to confirm its current co-orbital nature 
     with Venus. This paper is organized as follows: in Section 2, we focus on the available information on 2012 XE$_{133}$, summarize some 
     facts on co-orbital minor bodies and briefly outline our numerical model. Section 3 presents an analysis of the past, present and 
     future dynamical evolution of 2012 XE$_{133}$. Section 4 provides a comparative analysis that puts the object into context among the 
     other two previously known Venus co-orbitals. Our results are discussed in Section 5 emphasizing the possible existence of additional
     objects like 2012 XE$_{133}$. Our conclusions are summarized in Section 6.
%
%
     \begin{table*}
      \fontsize{8}{11pt}\selectfont
      \tabcolsep 0.35truecm
      \caption{Heliocentric Keplerian orbital elements of asteroids 2012 XE$_{133}$, (322756) 2001 CK$_{32}$ and 2002 VE$_{68}$. Values 
               include the 1$\sigma$ uncertainty. The orbit of 2012 XE$_{133}$ is based on 102 observations with a data-arc span of 28 
               d. Although the quality of the orbit of 2012 XE$_{133}$ is currently lower than that of the other two co-orbitals, it is
               comparable to the level of precision available when the other two objects were identified as orbital companions to Venus.  
               (Epoch = JD2456400.5, 2013-Apr-18.0; J2000.0 ecliptic and equinox. Source: JPL Small-Body Database.)
              }
      \begin{tabular}{ccccc}
       \hline
                                                               &   &                       &   (322756)                    &                                \\
                                                               &   &   2012 XE$_{133}$     &   2001 CK$_{32}$              &   2002 VE$_{68}$               \\
       \hline
        Semi-major axis, $a$ (au)                              & = &   0.72297$\pm$0.00014 &   0.725476965$\pm$0.000000004 &   0.7236506697$\pm$0.0000000005\\
        Eccentricity, $e$                                      & = &   0.4332$\pm$0.0003   &   0.3824157$\pm$0.0000002     &   0.41036031$\pm$0.00000005    \\
        Inclination, $i$ ($^{\circ}$)                          & = &   6.711$\pm$0.008     &   8.13191$\pm$0.00002         &   9.006018$\pm$0.000013        \\
        Longitude of the ascending node, $\Omega$ ($^{\circ}$) & = & 281.096$\pm$0.006     & 109.5025$\pm$0.0002           & 231.582435$\pm$0.000005        \\
        Argument of perihelion, $\omega$ ($^{\circ}$)          & = & 337.078$\pm$0.007     & 234.1055$\pm$0.0002           & 355.462382$\pm$0.000014        \\
        Mean anomaly, $M$ ($^{\circ}$)                         & = &  30.98$\pm$0.06       &  42.31827$\pm$0.00009         & 163.09449$\pm$0.00003          \\
        Perihelion, $q$ (au)                                   & = &   0.4098$\pm$0.0003   &   0.44804316$\pm$0.00000011   &   0.42669316$\pm$0.00000004    \\
        Aphelion, $Q$ (au)                                     & = &   1.0361$\pm$0.0002   &   1.002910769$\pm$0.000000006 &   1.0206081798$\pm$0.0000000007\\
        Absolute magnitude, $H$ (mag)                          & = &  23.4$\pm$0.6         &  18.7$\pm$1.0                 &  20.5$\pm$0.4                  \\
       \hline
      \end{tabular}
      \label{elements}
     \end{table*}
%
%
%
%
     \begin{figure}
       \centering
        \includegraphics[width=\linewidth]{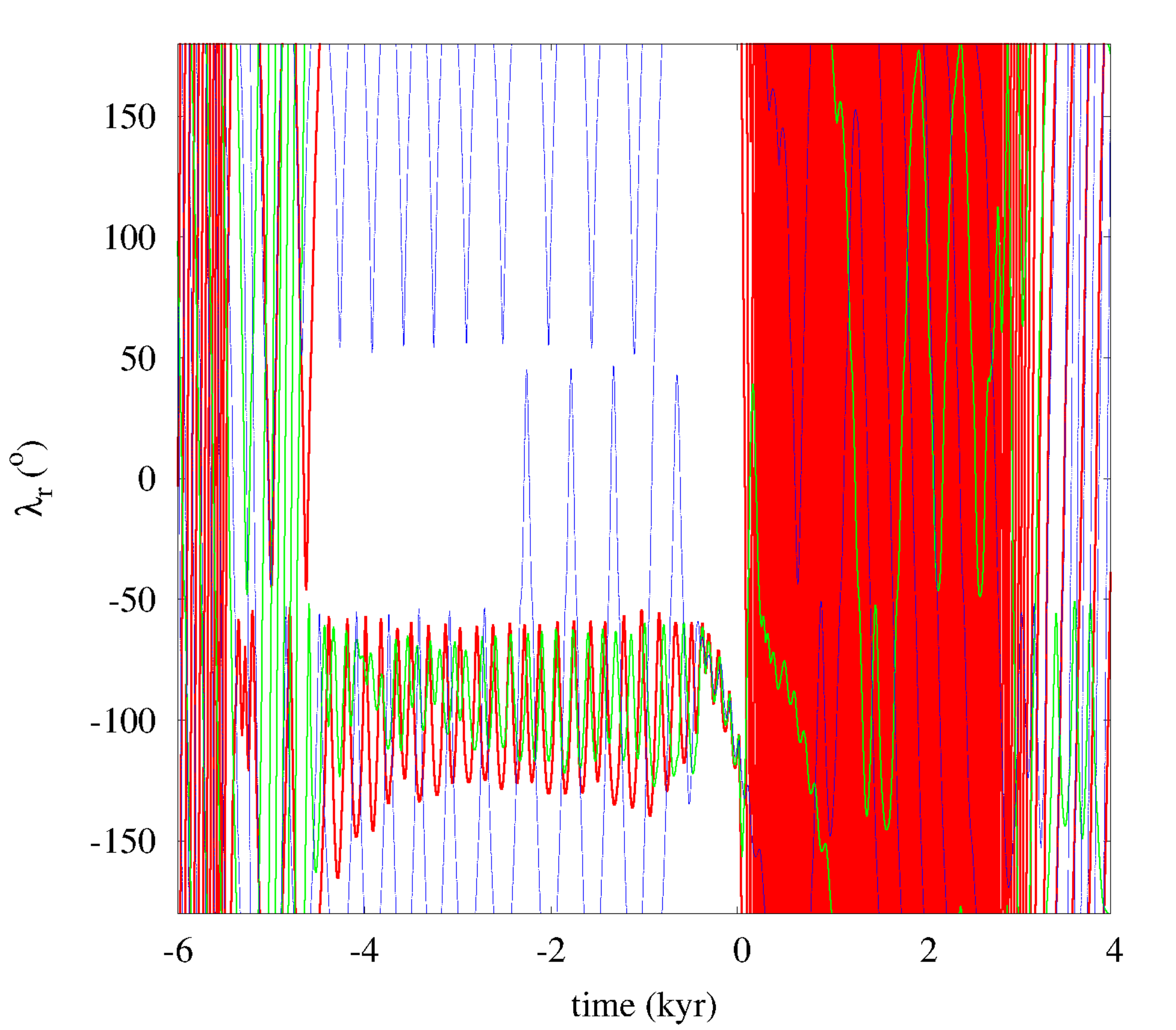}
        \caption{The resonant angle, $\lambda_{r}$ for the nominal orbit of 2012 XE$_{133}$ in Table \ref{elements} (thick red line) and two 
                 of the control orbits. The thinnest blue line corresponds to a particular control orbit that was chosen close to the 
                 3$\sigma$ limit so its orbital elements are most different from the nominal ones. The remaining orbit (green line) has 
                 osculating elements within 1$\sigma$ of those of the nominal orbit. The output time-step for these plots is 1.1 d and no 
                 filtering or smoothing was applied to the data (same for subsequent figures unless explicitly noted).
                }
        \label{lambda}
     \end{figure}
%
%
%
%
     \begin{figure}
       \centering
        \includegraphics[width=\linewidth]{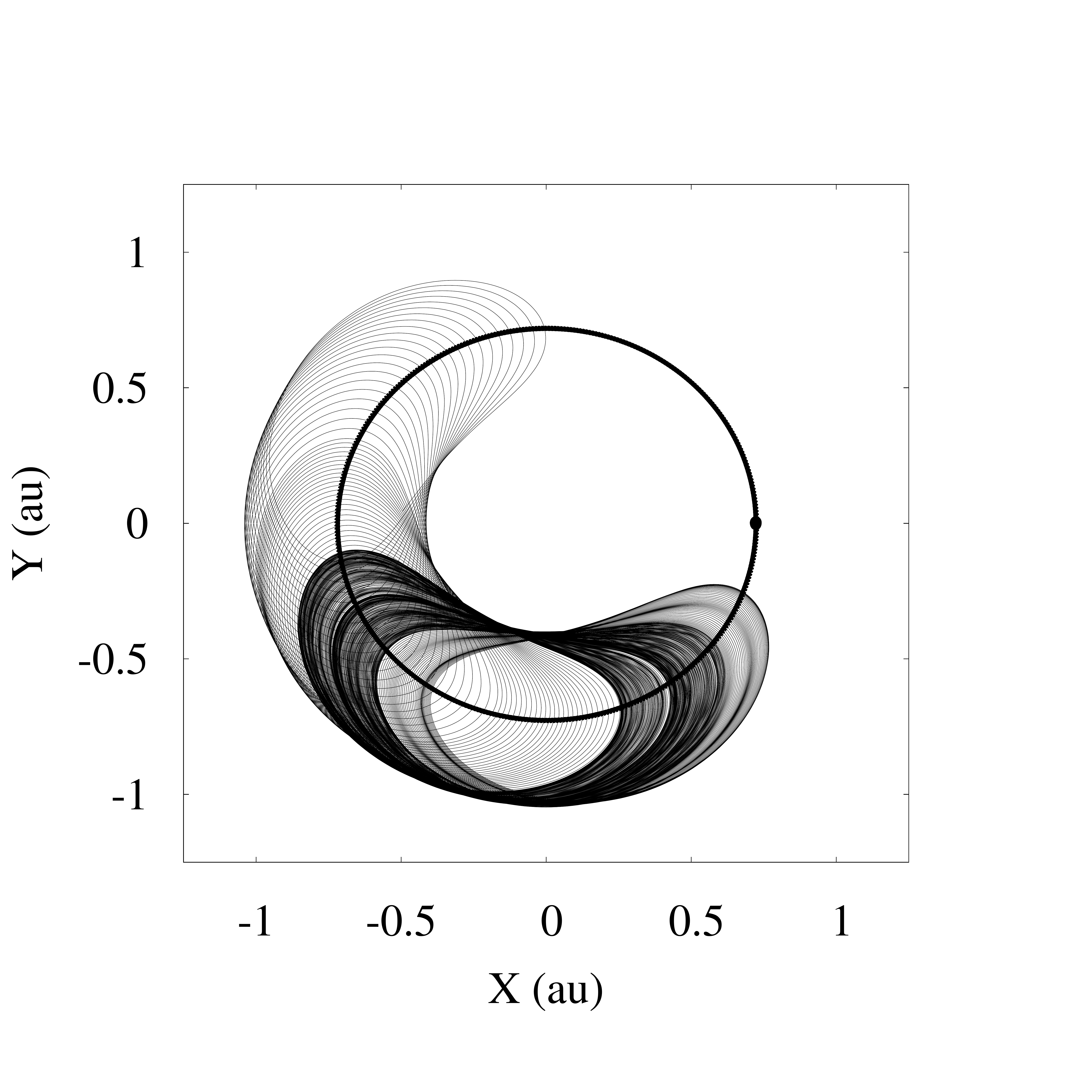}
        \caption{The motion of 2012 XE$_{133}$ over the time range (-200, 50) yr is displayed projected onto the ecliptic plane in a 
                 coordinate system rotating with Venus. The orbit and the position of Venus are also indicated. 
                }
        \label{hs}
     \end{figure}
%
%

  \section{Initial conditions and numerical model}
     Minor planet 2012 XE$_{133}$ was discovered by Jess A. Johnson working for the Catalina Sky Survey at Mt. Bigelow, Arizona on 2012 
     December 12. Its apparent magnitude was $V$ = 18.5. The object was observed on the same night by the Lunar and Planetary Laboratory 
     (LPL)/Spacewatch II project and from 
     the Magdalena Ridge Observatory, Socorro (McMillan et al. 2012)\footnote{http://www.minorplanetcenter.net/mpec/K12/K12X85.html}. With a 
     value of the semi-major axis $a$ = 0.7230 au, very close to that of Venus (0.7233 au), this Aten asteroid is an NEO moving in a quite 
     eccentric, $e$ = 0.43, and slightly inclined, $i = 6\fdg7$, orbit that makes it a Mercury grazer, Venus crosser, and Earth 
     crosser. Therefore, its orbit is similar to those of the two previously known Venus co-orbitals (see Table \ref{elements}): (322756) 2001 
     CK$_{32}$ and 2002 VE$_{68}$. Its current orbit is based on 102 observations with a data-arc span of 28 d. In comparison, the data-arc 
     spans of (322756) 2001 CK$_{32}$ and 2002 VE$_{68}$ were 512 d (with 82 observations) and 24 d (with 203 observations), respectively, 
     when they were identified as Venus co-orbitals by Brasser et al. (2004) and Mikkola et al. (2004). As a recent discovery, little is 
     known about this object besides its orbit with the exception of its absolute magnitude, $H$ = 23.4$\pm$0.6~mag (assumed $G$ = 0.15), 
     which indicates a diameter in the range 62-138 m for an assumed albedo in the range 0.20-0.04. One of the objectives of this work is to 
     encourage further observations of this small but interesting object.
     \hfil\par
     We selected 2012 XE$_{133}$ as a candidate Venus co-orbital because of its small relative semi-major axis, $|a - a_{Venus}| \sim$ 
     0.0004 au. In the Solar system, an object is co-orbital to a planet if their orbital periods and therefore their semi-major axes are 
     nearly equal. Once an object is trapped in a 1:1 mean motion resonance and depending on its relative energy with respect to the host 
     planet (H\'enon 1969), it can describe any of the three main orbit types: quasi-satellite, tadpole or horseshoe. The names refer to the 
     actual shapes of the orbits when seen in a frame of reference rotating with the host planet. The key parameter to differentiate among 
     these three main orbit types and additional compound states, as first described by Namouni (1999), is the libration of the difference 
     between the mean longitudes of the object and its host planet or relative mean longitude, $\lambda_{r}$. The mean longitude of an 
     object is given by $\lambda$ = $M$ + $\Omega$ + $\omega$, where $M$ is the mean anomaly, $\Omega$ is the longitude of ascending node 
     and $\omega$ is the argument of perihelion. If the object follows a quasi-satellite orbit, $\lambda_{r}$ oscillates around 0$^{\circ}$. 
     When the object moves in a tadpole orbit, $\lambda_{r}$ librates around 60$^{\circ}$ (L$_4$ trojan) or -60$^{\circ}$ (or 300$^{\circ}$, 
     L$_5$ trojan). For objects moving in eccentric and/or inclined orbits the location of the Lagrangian points L$_4$ and L$_5$ changes; 
     for example, in an orbit as eccentric as that of 2012 XE$_{133}$, it is shifted in $\lambda$ towards the L$_3$ point by 
     30$^{\circ}$-40$^{\circ}$ (Namouni \& Murray 2000). If the libration amplitude is larger than 180$^{\circ}$, encompassing L$_3$, L$_4$ 
     and L$_5$ but not reaching the actual planet, it is said that the object follows a symmetric horseshoe orbit. These are the three 
     principal dynamical states included in general co-orbital motion (see, e.g., Murray \& Dermott 1999 and Mikkola et al. 2006 for 
     additional details). Compound states are also possible in which the object may librate around 0$^{\circ}$ with an amplitude $>$ 
     180$^{\circ}$ encompassing L$_4$ and L$_5$ (compound quasi-satellite-tadpole orbit), asymmetric horseshoe orbits 
     (horseshoe-quasi-satellite orbiters) in which the libration amplitude $>$ 270$^{\circ}$, encompassing the planet, and a few other 
     combinations (see, e.g., Namouni 1999; Namouni, Christou \& Murray 1999). When the relative mean longitude oscillates freely, we say 
     that the object is no longer in a 1:1 mean motion resonance with the planet, i.e. it becomes a passing object. 
     \hfil\par
     In order to confirm the co-orbital nature of 2012 XE$_{133}$ with Venus, we have performed $N$-body calculations using the Hermite 
     integrator (Makino 1991; Aarseth 2003) in both directions of time for 10 kyr. Our model Solar system includes the perturbations by the 
     eight major planets and treat the Earth and the Moon as two separate objects. It also includes the barycentre of the dwarf planet
     Pluto-Charon system and the three largest asteroids. The standard version of this direct $N$-body code is publicly available from the
     IoA web site\footnote{http://www.ast.cam.ac.uk/$\sim$sverre/web/pages/nbody.htm}. Our integrations of the full equations of motion
     neglect relativistic terms as point (constant) mass objects moving in a conservative system are assumed. The role of the Yarkovsky and
     Yarkovsky-O'Keefe-Radzievskii-Paddack (YORP) effects (see, e.g., Bottke et al. 2006) is also ignored. We use initial conditions
     (positions and velocities in the barycentre of the Solar system referred to the JD2456400.5 epoch) provided by the JPL's
     \textsc{HORIZONS} ephemeris system (Giorgini et al. 1996; Standish 1998). In all the figures, $t$ = 0 coincides with the JD2456400.5
     epoch. Relative errors in the total energy at the end of the simulations are $< 1 \times 10^{-15}$. In addition to the calculations
     completed using the nominal orbital elements in Table \ref{elements} for 2012 XE$_{133}$, we have performed 50 control simulations
     using sets of orbital elements obtained from the nominal ones within the accepted uncertainties. The computed set of control orbits
     follows a normal distribution in the six-dimensional space of orbital elements. These synthetic orbital elements are compatible with 
     the nominal ones within the 3$\sigma$ deviations (see Table \ref{elements}) and reflect the observational uncertainties in astrometry.
     The results of these calculations clearly show (see Section 3 and Fig. \ref{lambda}) that the true phase-space trajectory will diverge 
     exponentially from that obtained from the nominal orbital elements in Table \ref{elements} within a very short time-scale. This object 
     and the other two Venus co-orbitals have e-folding times of the order of 100 yr. The source of the Heliocentric Keplerian osculating 
     orbital elements and uncertainties is the JPL Small-Body Database\footnote{http://ssd.jpl.nasa.gov/sbdb.cgi}. Additional details can be 
     found in de la Fuente Marcos \& de la Fuente Marcos (2012).
%
%
     \begin{figure*}
        \centering
        \includemedia[
          label=2012XE133,
          width=\textwidth,height=0.65\textwidth,
          activate=onclick,
          addresource=2012XE133w.mp4,
          flashvars={
            source=2012XE133w.mp4
           &autoPlay=true
           &loop=true
           &controlBarMode=floating
           &controlBarAutoHide=false
           &scaleMode=letterbox
          }
        ]{\includegraphics{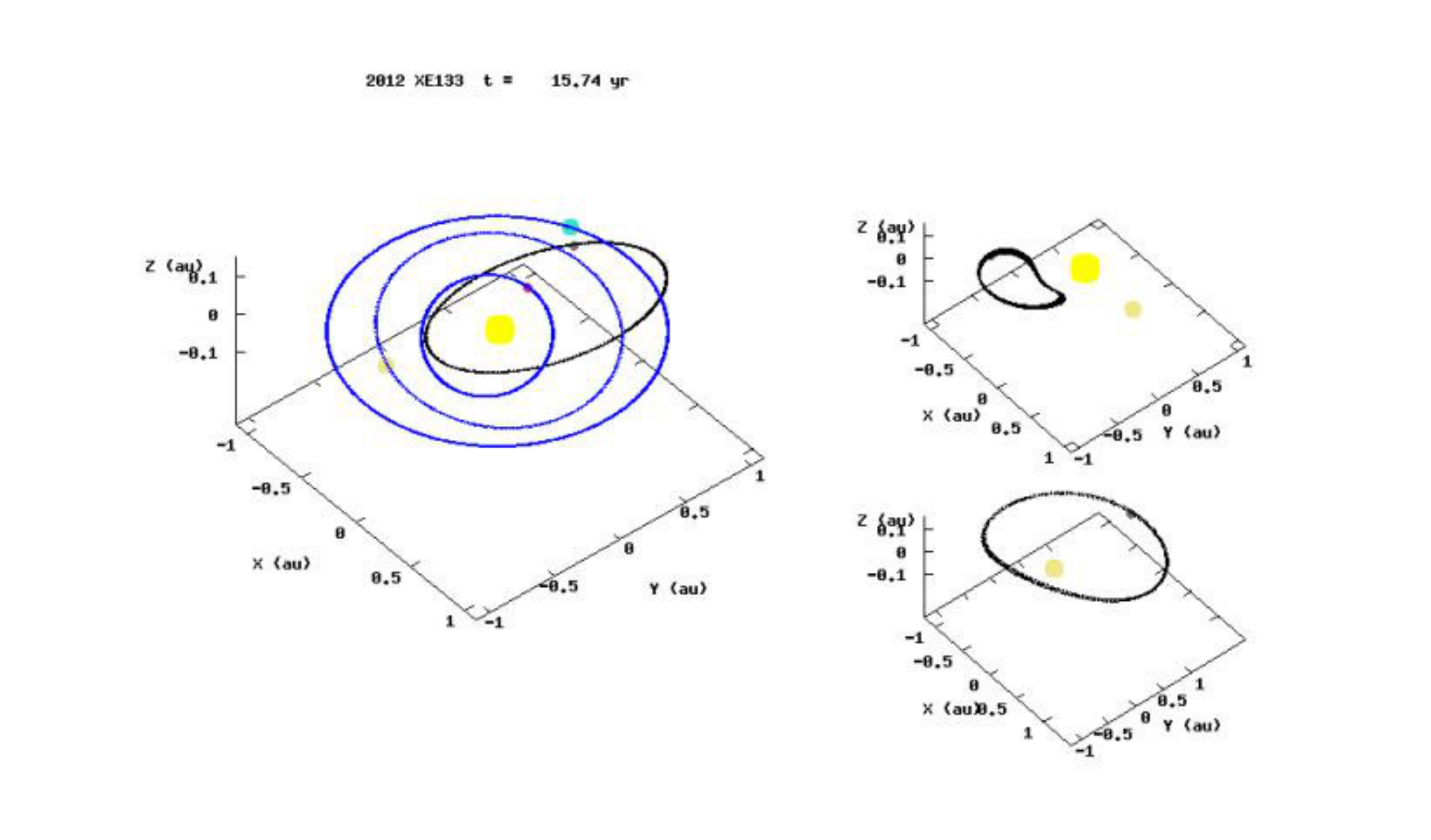}}{VPlayer.swf}
        \PushButton[
           onclick={
             annotRM['2012XE133'].activated=true;
             annotRM['2012XE133'].callAS('play');
           }
        ]{\fbox{Play}}
        \PushButton[
           onclick={
             annotRM['2012XE133'].activated=true;
             annotRM['2012XE133'].callAS('pause');
           }
        ]{\fbox{Pause}}
        \caption{Three-dimensional evolution of the orbit of 2012 XE$_{133}$ for a few decades starting on JD2456400.5, 2013-Apr-18.0, in 
                 three different frames of reference: heliocentric (left), frame corotating with Venus but centred on the Sun (top-right), 
                 and centred on Venus (bottom-right). The grey point denotes 2012 XE$_{133}$, the pink one denotes Mercury, the khaki one denotes Venus, 
                 the turquoise one denotes the Earth and the Sun appears in yellow. The osculating orbits are outlined and the viewing angle 
                 changes slowly to facilitate visualizing the overall orbital evolution. The current dynamical state resembles a tadpole 
                 orbit even if 2012 XE$_{133}$ already left the Lagrangian point L$_5$ after a close encounter with the Earth nearly 350 yr
                 ago.} 
        \label{animation}
     \end{figure*}
%
%

  \section{Orbital evolution: past, present and future}
\begin{figure*}
       \centering
        \includegraphics[width=\linewidth]{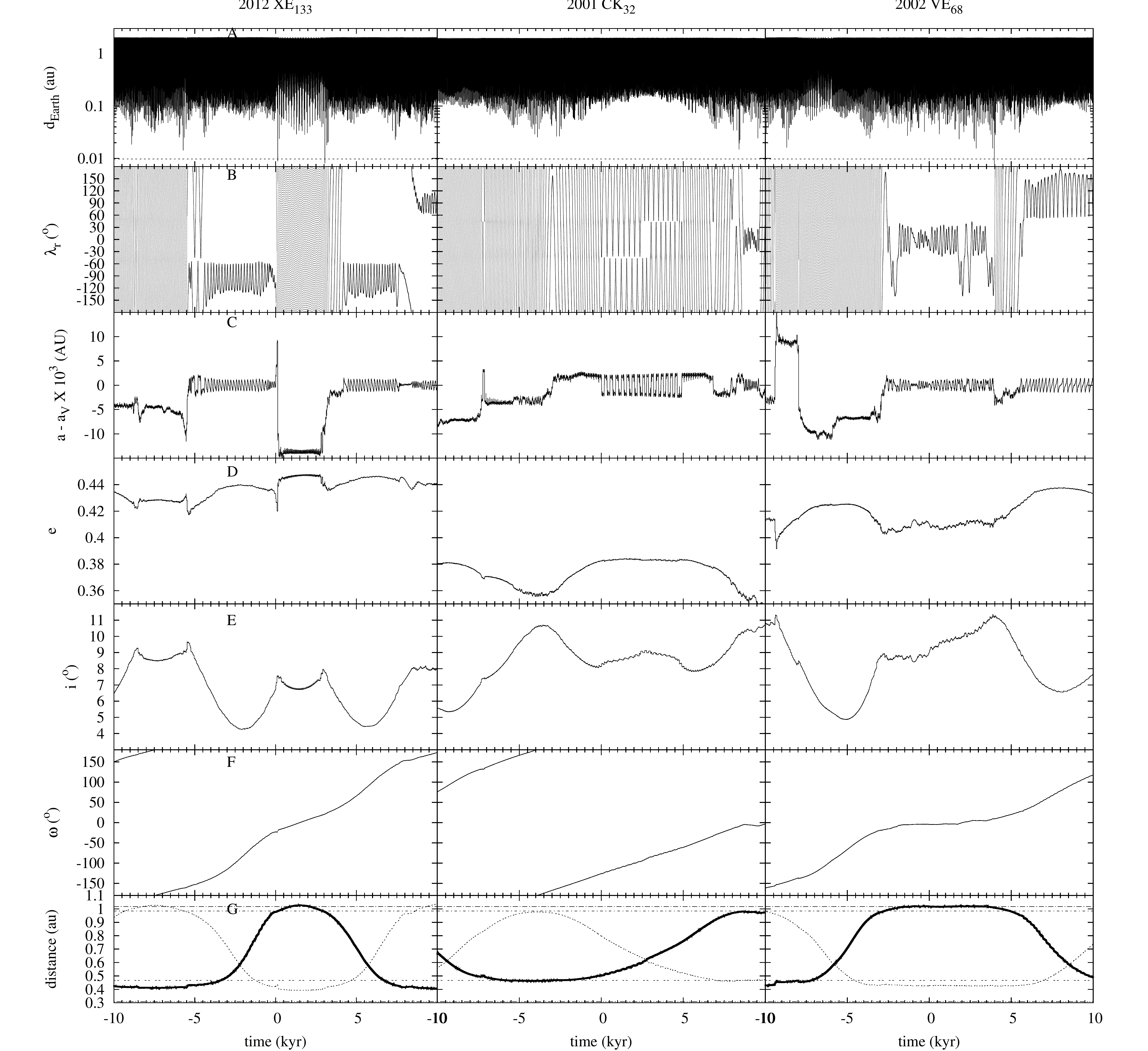}
        \caption{Comparative time evolution of various parameters for the three known Venus co-orbitals: 2012 XE$_{133}$ (left), (322756) 
                 2001 CK$_{32}$ (center) and 2002 VE$_{68}$ (right). The distance from the Earth (panel A); the value of the Hill sphere 
                 radius of the Earth, 0.0098 au, is displayed. The resonant angle, $\lambda_{r}$ (panel B) for the nominal orbit in Table 
                 \ref{elements}. The orbital elements $a - a_{\rm Venus}$ (panel C), $e$ (panel D), $i$ (panel E) and $\omega$ (panel F). The 
                 distance to the descending (thick line) and ascending nodes (dotted line) appears in panel G. Earth's and Mercury's 
                 aphelion and perihelion distances are also shown. The distance to the descending node of 2012 XE$_{133}$ currently 
                 coincides with Earth's perihelion, that of (322756) 2001 CK$_{32}$ is close to Mercury's aphelion and 2002 VE$_{68}$'s 
                 matches Earth's aphelion. The output time-step for these plots is 0.3 yr.
                }
        \label{all}
     \end{figure*}

\begin{figure}
       \centering
        \includegraphics[width=\linewidth]{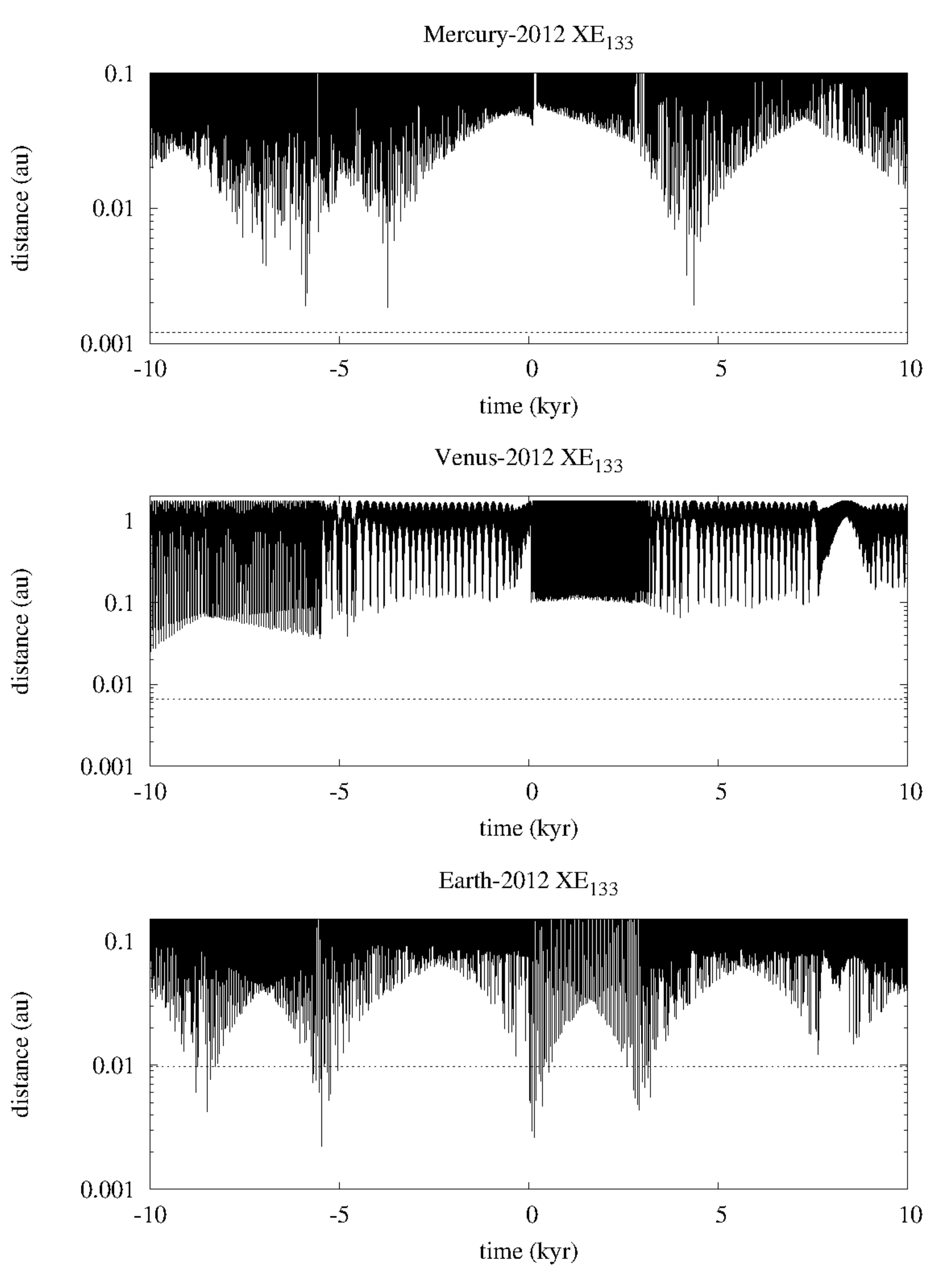}
        \caption{The distance of 2012 XE$_{133}$ from Mercury (top panel), Venus (middle panel), and the Earth (bottom panel). Close 
                 encounters are possible both at perihelion (with Mercury) and at aphelion (with the Earth). The values of the Hill sphere
                 radii are also displayed. The output time-step for these plots is 1.1 d.
                }
        \label{distances}
     \end{figure}

     The motion of 2012 XE$_{133}$ over the time range (-200, 50) yr as seen in a coordinate system rotating with Venus projected onto the 
     ecliptic plane is plotted in Fig. \ref{hs} and the three-dimensional evolution of its orbit in multiple frames of reference over the 
     next few decades is displayed as an animation in Fig. \ref{animation}. The object is a Venus co-orbital currently following a 
     transitional trajectory between Venus' Lagrangian points L$_5$ and L$_3$. Our calculations (see Figs \ref{lambda} and \ref{all}, left-hand 
     panels) show that 2012 XE$_{133}$ has already remained in the 1:1 commensurability with Venus for several thousand years. For most of 
     that time (4000 yr), it has been following a tadpole orbit in the neighbourhood of Venus' L$_5$ point. After a close encounter with the
     Earth at 0.012 au 346 yr ago, it started moving away from the Lagrangian point L$_5$ to undergo a transition between its previous 
     tadpole orbit and a horseshoe orbit. Horseshoe orbits are generally less stable than tadpole orbits, and they are often considered as a 
     transitional state before ejection from a 1:1 resonance (see, e.g., Murray \& Dermott 1999). However, for the particular case of 
     Earth's and Venus' horseshoe librators, long-lived configurations are possible (\'Cuk, Hamilton \& Holman 2012). All the computed 
     control orbits confirm that 2012 XE$_{133}$ currently follows a rather irregular path of the horseshoe type, approaching the 
     neighbourhood of L$_3$. The current precession rate of the bean-shaped loops associated with the asteroid orbital period is very slow and 
     this transitional path will last nearly 150-300 yr. Then the object's resonant angle will circulate after leaving the 1:1 resonance 
     with Venus as a result of a very close encounter with the Earth at about 0.003 au in 2098 CE. Its future orbital evolution depends 
     strongly on the outcome of a very close encounter with the Earth that will take place in the time frame 85-340 yr from now. Nearly half of 
     the control orbits evolve into passing orbits beyond that point, to never return to the Venus co-orbital region. The rest remain in or 
     near the co-orbital region for thousands of years, at least $\sim$3000 yr. For those orbits, a variety of dynamical behaviours are 
     observed, including compound quasi-satellite-tadpole episodes and L$_5$ tadpole orbits but also short circulation events. As for the 
     past, nearly 70 per cent of the control orbits remained as an L$_5$ trojan for about 3-5 kyr prior to the close encounter with the Earth that 
     sends the object in the neighbourhood of L$_3$. The remaining control orbits exhibit various horseshoe orbit types including not only symmetric 
     but also asymmetric ones. Beyond about 3000 yr into the past, the orbit is difficult to predict. Nearly 10 per cent of the control orbits 
     remain within the Venus co-orbital region for the entire simulated time, 20 kyr, but alternating multiple dynamical states and 
     circulation episodes. Repetitive episodes as the ones described here, in which the relative mean longitude librates for several cycles and 
     then circulates for a few more cycles before restarting libration once again, are characteristic of a type of dynamical behaviour known 
     as resonance angle nodding, see Ketchum, Adams \& Bloch (2013). These authors concluded that nodding often occurs when a small body is in an 
     external (near) mean motion resonance with a larger planet. This type of complicated dynamics has been observed in other horseshoe 
     librators. During the studied timespan, 2012 XE$_{133}$ remains at a safe distance from Venus (see Fig. \ref{distances}), but close
     encounters are possible both at perihelion (with Mercury) and at aphelion (with the Earth); these encounters are critical in changing
     the dynamical status of the object as discussed in the following section.  
%
%
%
%
%
%
%
%
  \section{Asteroid 2012 XE$_{133}$ in context}
\begin{figure*}
        \centering
        \includemedia[
          label=known1,
          width=\textwidth,height=0.5\textwidth,
          activate=onclick,
          addresource=2001CK32w.mp4,
          flashvars={
            source=2001CK32w.mp4
           &autoPlay=true
           &loop=true
           &controlBarMode=floating
           &controlBarAutoHide=false
           &scaleMode=letterbox
          }
        ]{\includegraphics{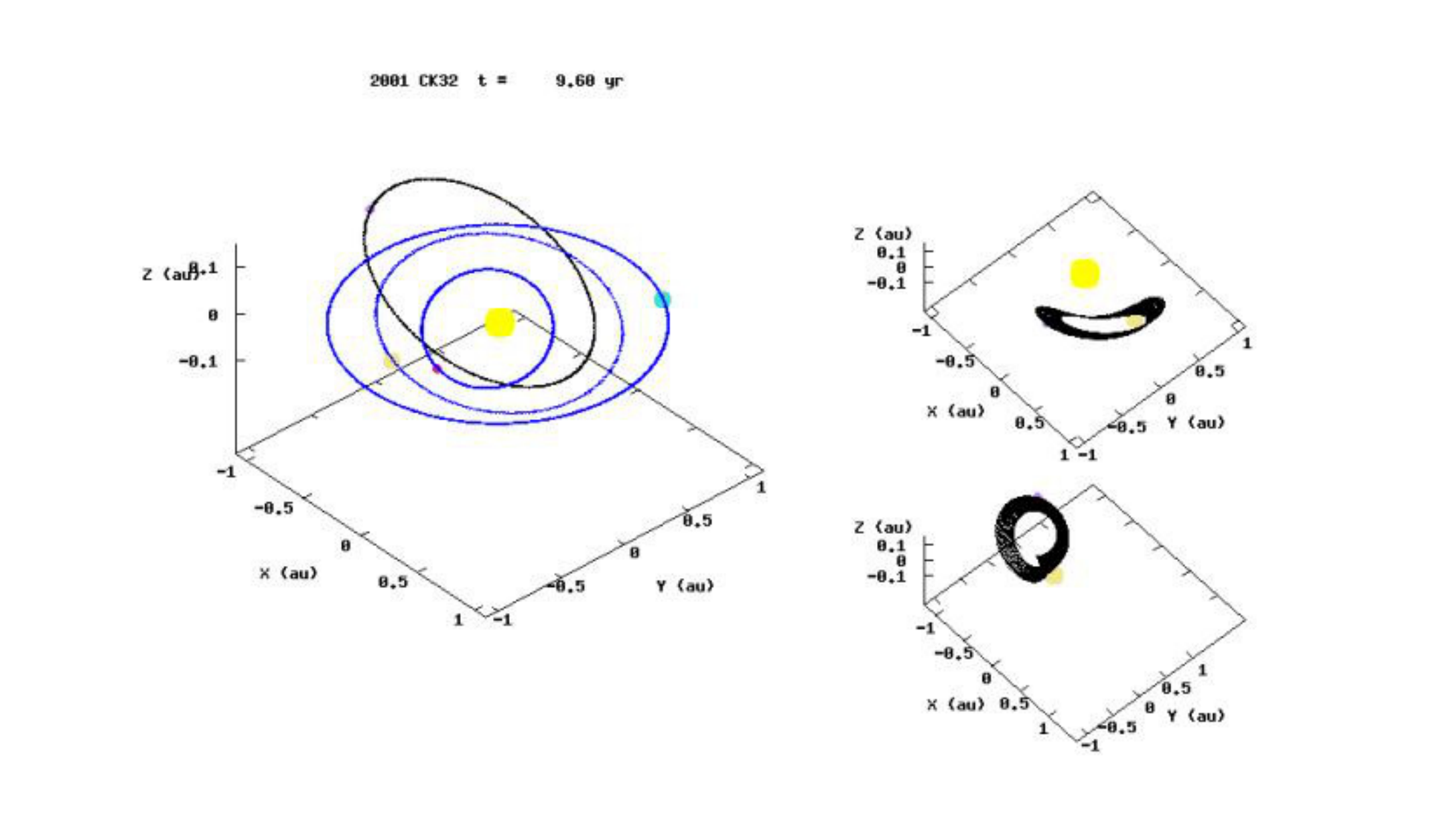}}{VPlayer.swf}
        \includemedia[
          label=known2,
          width=\textwidth,height=0.5\textwidth,
          activate=onclick,
          addresource=2002VE68w.mp4,
          flashvars={
            source=2002VE68w.mp4
           &autoPlay=true
           &loop=true
           &controlBarMode=floating
           &controlBarAutoHide=false
           &scaleMode=letterbox
          }
        ]{\includegraphics{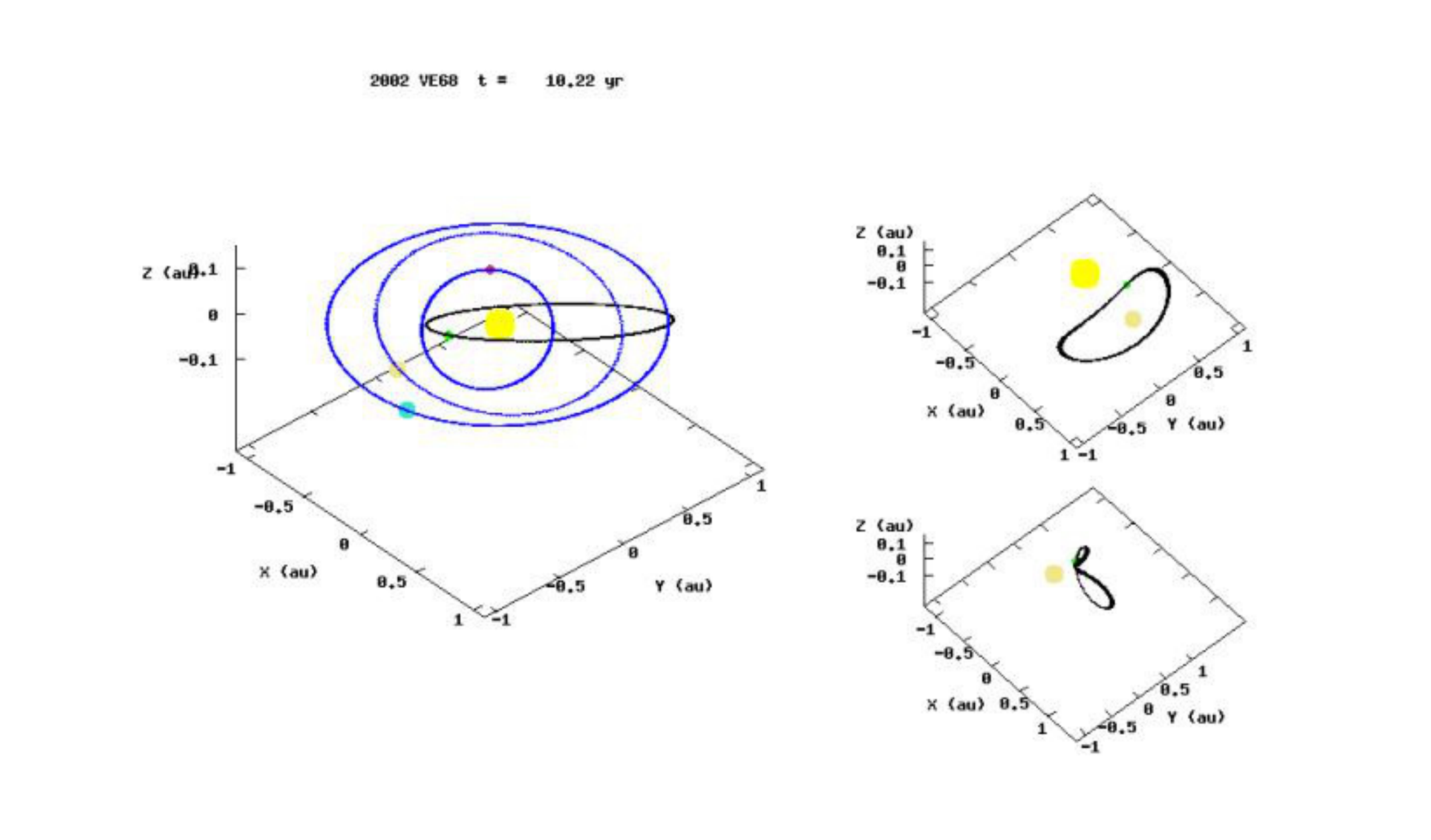}}{VPlayer.swf}
        \PushButton[
           onclick={
             annotRM['known1'].activated=true;
             annotRM['known1'].callAS('playPause');
             annotRM['known2'].activated=true;
             annotRM['known2'].callAS('playPause');
           }
        ]{\fbox{Play/Pause}}
        \caption{Short term three-dimensional evolution of the orbit of (322756) 2001 CK$_{32}$ (top) and 2002 VE$_{68}$ (bottom) in three 
                 different frames of reference: heliocentric (left), frame corotating with Venus but centred on the Sun (top-right) and 
                 centred on Venus (bottom-right). The purple point indicates 2012 XE$_{133}$, the lime point indicates 2002 VE$_{68}$, the pink one indicates 
                 Mercury, the khaki one indicates Venus, the turquoise one indicates the Earth and the Sun appears in yellow. The osculating orbits are 
                 outlined and the viewing angle changes slowly to facilitate visualizing the overall orbital evolution. Asteroid 322756 is a 
                 horseshoe-quasi-satellite orbiter that currently follows the quasi-satellite stage of the compound orbit. 2002 VE$_{68}$ 
                 exhibits typical quasi-satellite behaviour.} 
        \label{animation2}
     \end{figure*}

     Fig. \ref{all} displays the comparative evolution of the osculating orbital elements of all the known Venus co-orbitals (nominal 
     orbits in Table \ref{elements}). The short term three-dimensional evolution of the orbit of (322756) 2001 CK$_{32}$ and 2002 VE$_{68}$ 
     is displayed as an animation in Fig. \ref{animation2}. There are many similarities but also obvious differences, mostly in terms 
     of short-term stability. All of them move in rather eccentric orbits with moderate inclinations and all of them are Mercury grazers, 
     Venus crossers and Earth crossers. As the orbits of 322756, 2002 VE$_{68}$ and 2012 XE$_{133}$ are very much alike, the overall 
     details of the mechanism that controls their orbital behaviour must be similar too. When moving in a 1:1 mean motion resonance, 
     transfers between quasi-satellite, horseshoe and tadpole orbits are the result of the libration of the nodes (Wiegert, Innanen \& 
     Mikkola 1998). Our analysis of the dynamics of 2002 VE$_{68}$ in de la Fuente Marcos \& de la Fuente Marcos (2012) confirmed the result 
     obtained by Mikkola et al. (2004): during the quasi-satellite phase the distance to the descending node of the object remains 
     remarkably close to the value of Earth's aphelion and the distance to the ascending node is also relatively close to Mercury's aphelion. 
     Here we have a similar situation. In the Solar system and for an object following an inclined path, close encounters with major planets 
     are only possible in the vicinity of the nodes. The distance between the Sun and the nodes is given by $r = a (1 - e^2) / (1 \pm e \cos 
     \omega)$, where the '+' sign is for the ascending node and the '-' sign is for the descending node. Close encounters at the nodes 
     control the dynamics of 2002 VE$_{68}$ and Fig. \ref{all}, panels A, B and G, clearly reveals that it is also the case for 322756 and 
     2012 XE$_{133}$. In particular, the future evolution of the distances to the nodes of 2002 VE$_{68}$ and 2012 XE$_{133}$ is analogous. 
     We can therefore conclude that 322756, 2002 VE$_{68}$ and the recently discovered 2012 XE$_{133}$ are part of a dynamical group which 
     is characterized by the libration of their nodes between the orbits of Mercury and the Earth. As a result, the gravitational 
     perturbations from the Earth and Mercury are currently most effective in keeping these asteroids at a safe distance from Venus. At 
     present, 2012 XE$_{133}$ approaches the Earth in the vicinity of its descending node in November-December every eighth year; in a way, 
     this contributes to stabilizing the orbit if the distance of the closest approach is larger than about 0.03 au. These three objects orbit the 
     Sun in a near 8:13 resonance with the Earth but they also move in a near 9:23 resonance with Mercury although this is less clear than 
     the near resonance with the Earth. The three objects exhibit resonant (or near resonant) behaviour with Mercury, Venus and the Earth. 
     However, 2012 XE$_{133}$ follows the most eccentric orbit and has the smallest inclination, which translates into more frequent and 
     closer encounters with the Earth and explains why its orbit is the most unstable of the three. Encounters with Mercury as close as 
     0.0011 au and with Venus (only in the future) at 0.002 au have also been observed in a few of the control orbits. Close encounters with 
     Venus are only observed when the nodes are far from the perihelion/aphelion positions.
     \hfil\par
     Our results for 322756 are fully compatible with those in Brasser et al. (2004, fig. 1) for the next 2000 yr but differences appear
     beyond that point. These arise from the fact that we are using the newest orbital elements, different physical models and the intrinsic 
     chaos that drives the dynamical evolution of the object. 
%
%
%
%

  \section{Discussion}
\begin{figure}
       \centering
        \includegraphics[width=\linewidth]{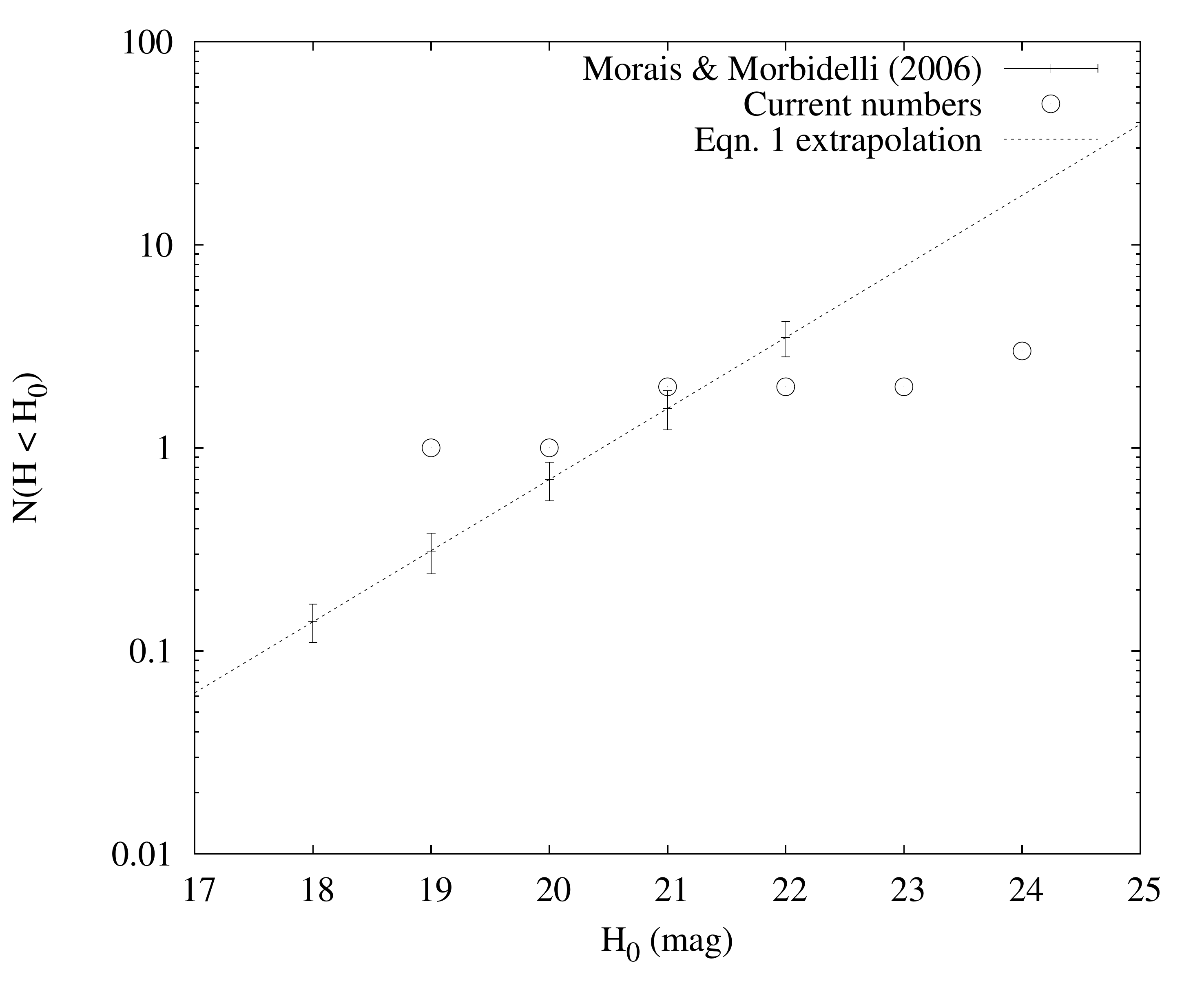}
        \caption{The number distribution of Venus co-orbitals as a function of the absolute magnitude. This plot displays results obtained
                 by Morais \& Morbidelli (2006) with error bars, the current value of the counts (including 2012 XE$_{133}$) and the
                 predictions from equation (\ref{law}).
                }
        \label{size}
     \end{figure}

     Morais \& Morbidelli (2006) studied the population of Venus co-orbitals that are also NEOs and concluded that the average duration of a 
     Venus co-orbital episode is $\sim$32\,000 yr. Our calculations indicate that 2012 XE$_{133}$ current co-orbital episode is going to be 
     significantly shorter but it is also true that its absolute magnitude is higher than the limiting value in their study and, therefore, 
     their results may not apply in this case. They also estimated that the number of objects with absolute magnitude $< H_o$ that become 
     co-orbital with Venus over a time interval $\Delta t$ is $N \Delta t/\Delta t_{1:1}$, where $N$ is given in table 1 of their 2006 paper 
     and $\Delta t_{1:1}$ is the average duration of a Venus co-orbital episode. The model in Morais \& Morbidelli (2006) is based on a 
     previous model by Bottke et al. (2000, 2002) which only applies in the size range corresponding to $H < 22$. This limit (about 400 m in 
     diameter) reflects the transition from tensile strength dominated bodies to rubble piles or gravity-dominated aggregates (see, e.g., 
     Love \& Ahrens 1996). Even with this limitation in mind, it could be useful to have some estimate of the expected number of smaller 
     objects. If we assume a power law of the form $N(H) \propto 10^{\alpha H}$ and use the results obtained by Morais \& Morbidelli (2006) 
     to find the constant and the index $\alpha$, we obtain 
     \begin{equation}
        N(H < H_o) = 6.98\times10^{-8} \ 10^{0.35\ H} \,, \label{law}
     \end{equation}
     where $N(H < H_o)$ is the number of objects with absolute magnitude $< H_o$. This simple extrapolation (plotted in Fig. \ref{size}) 
     predicts about 14 objects for $H < 24$, which translates into about 10 objects similar to 2012 XE$_{133}$ waiting to be discovered. But 
     if the average duration of co-orbital episodes is shorter for smaller, less massive objects as the case of 2012 XE$_{133}$ appears to 
     suggest, then the actual number of very small objects in co-orbital motion with Venus over a given timespan could be much larger in 
     relative terms than the equivalent figure for larger objects like 322756 or 2002 VE$_{68}$. It may be argued that because our present 
     calculations neglect the effects of non-gravitational forces like the Yarkovsky force, we cannot discuss the possibility of 
     size-dependent trends in our results but nominal orbits used to compute the initial conditions for our integrations are based on 
     the real positions of the objects, affected or not by non-gravitational forces. Therefore, the present orbital state of any observed 
     object is always the result of all the (real) forces acting on the object. If the Yarkovsky force has any effects on the orbital 
     evolution of the objects discussed here, those effects are already present on the initial conditions of the simulations but they are 
     gradually lost if the integrations do not account for the Yarkovsky force, which is the case here. The time-scale in which such 
     information is lost is likely longer than the characteristic e-folding time for these objects.
     \hfil\par
%
%
%
%
     The co-orbital identified in this paper was initially selected because of its small relative semi-major axis that is nearly 0.0004 au. 
     The relative semi-major axes of 322756 and 2002 VE$_{68}$ are 0.0021 au and 0.00032 au, respectively. During the last two decades, wide-field 
     NEO surveys have found thousands of objects passing in the neighbourhood of the Earth: could it be possible that some of these objects 
     may have not been properly identified and they are, in fact, trapped (even if temporarily) in a 1:1 commensurability with Venus? In 
     order to explore this possibility and try to uncover neglected Venus co-orbitals, we have studied the evolution of all the known 
     objects with relative semi-major axis (to Venus) $<$0.006 au. In this way, we have confirmed that no other known object (as of 2013 March 
     7) is currently co-orbital to Venus. However, we have found two additional objects, 2007 AG and 2002 LT$_ {24}$, that may 
     experience very brief co-orbital episodes (duration $<$10$^3$ yr) in the timespan $\pm$10$^4$ yr. The asteroid 2007 AG has an absolute
     magnitude of 20.13 (it is brighter than 2002 VE$_{68}$), a relative semi-major axis of 0.0030 au, an eccentricity of 0.37 and an inclination 
     of 11$\fdg$94. Its tadpole (L$_5$) and horseshoe episodes are few and brief ($<$ 900 yr); its orbit is well known with a data-arc 
     span of 1791 d. More interesting is the case of 2002 LT$_ {24}$ at $H$ = 21.85 with a relative semi-major axis of 0.0035 au, an eccentricity of 
     0.5 and an inclination of just 0$\fdg$76. It also experiences horseshoe behaviour with respect to Venus in several occasions, the 
     longest centred about 2000 yr into the future and lasting nearly 1000 yr. The orbit of this object is also very reliable with a 
     data-arc span of 2934 d. Like 2007 AG, 2002 LT$_ {24}$ is an Aten and NEO that is also included in the Potentially Hazardous Asteroid 
     (PHA) list. Finally, 2003 KO$_2$ appears to signal the edge of the co-rotational region of Venus at a relative semi-major axis of 
     0.0041 au. Again, an Aten, NEO and PHA with $H$ = 20.11, an eccentricity of 0.5 and an inclination of 23$\fdg$5, it experiences a 
     single half horseshoe loop during the studied timespan. All the objects move in rather eccentric orbits. In summary and although the 
     model in Morais \& Morbidelli (2006) merely provides statistical expectations, currently available data are compatible with predictions 
     in Fig. \ref{size}.
     \hfil\par
\begin{figure}
       \centering
        \includegraphics[width=\linewidth]{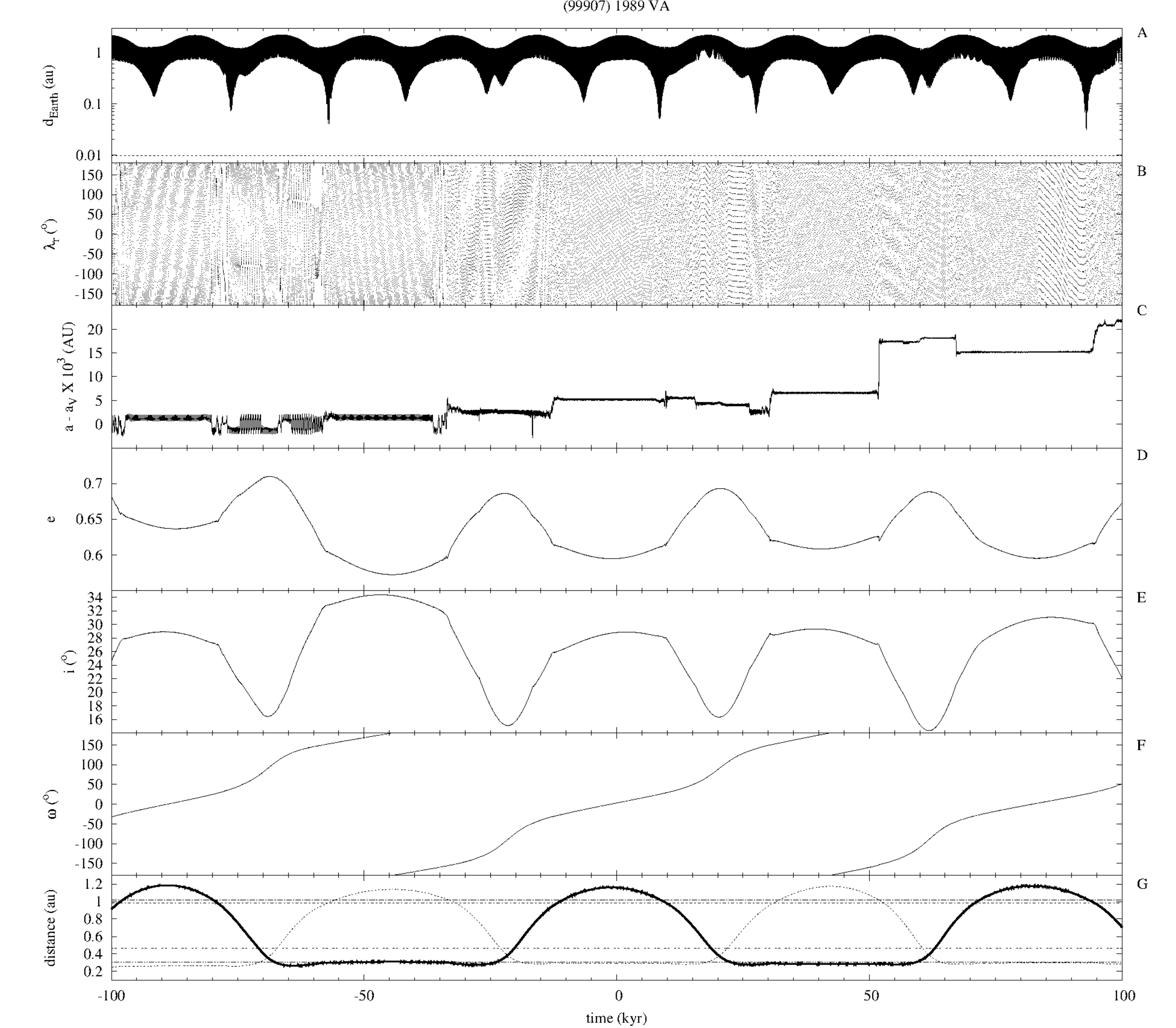}
        \caption{Same as Fig. \ref{all} but for (99907) 1989 VA. The nodes match Mercury's perihelion periodically and for extended periods 
                 of time. This minor body likely was a Venus co-orbital in the past.}
        \label{1989VA}
     \end{figure}
\begin{figure}
       \centering
        \includegraphics[width=\linewidth]{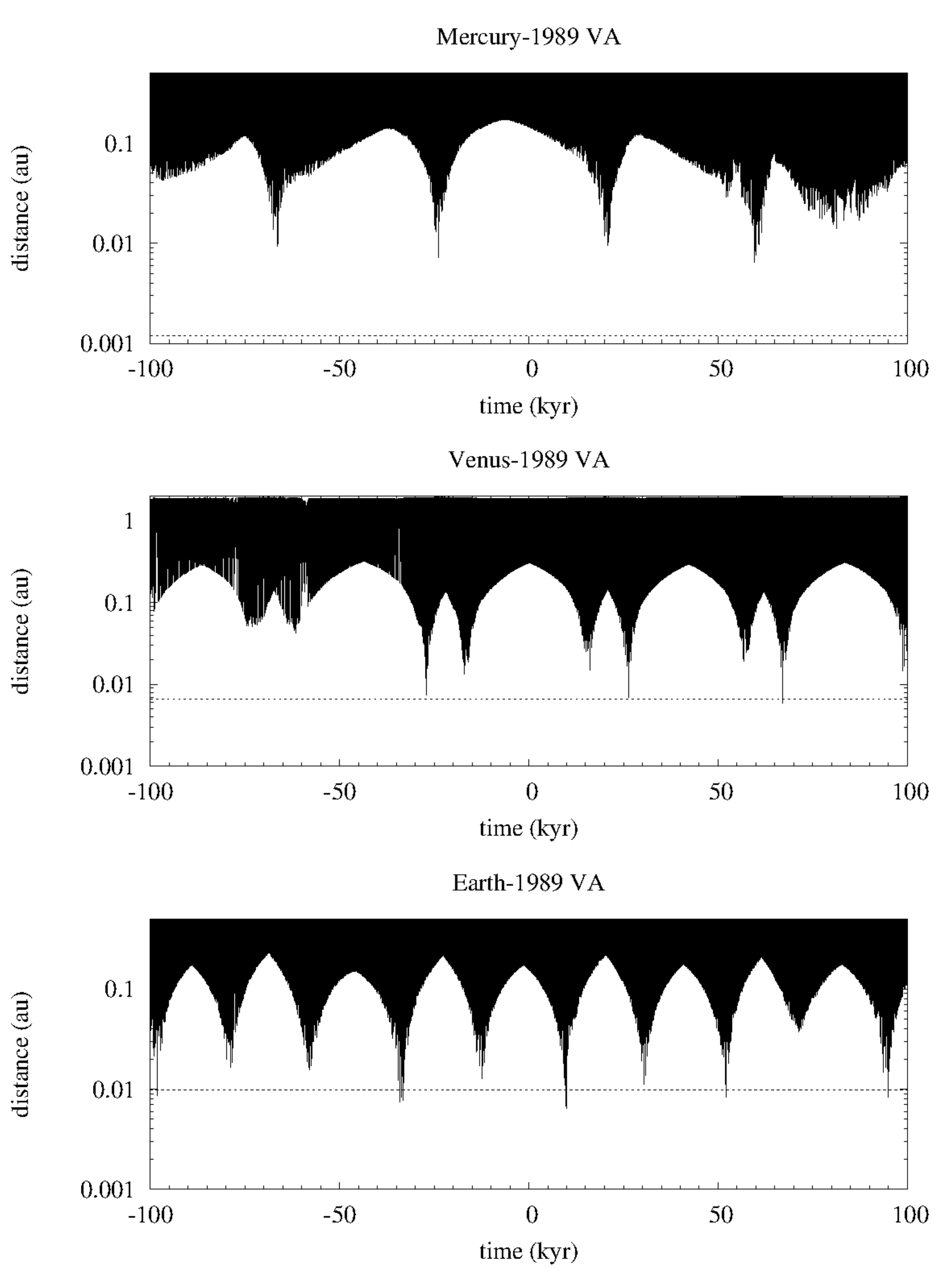}
        \caption{Same as Fig. \ref{distances} but for 1989 VA.
                }
        \label{distancesVA}
     \end{figure}

     Outside the timespan $\pm$10$^4$ yr, we confirm the early result obtained by Namouni et al. (1999) and Christou (2000) and pointed out 
     above, that the asteroid 99907 (relative semi-major axis 0.0051 au) was a Venus co-orbital. Christou (2000) found that in the 
     $\pm$10$^5$ yr interval this object has a 50 per cent probability of following co-orbital motion with Venus; we find that the object had 
     multiple co-orbital episodes in the past, some of them lasting thousands of years. However, its dynamics (see Fig. \ref{1989VA}) is 
     somewhat different from that of the objects previously discussed as it moves in a more eccentric (0.59) and inclined path 
     (28$\fdg$8). Now the distance of one of the nodal points of 99907 from the Sun usually coincides with the minimum distance to 
     Mercury from the Sun. Every 40 kyr both nodes reach Mercury's neighbourhood; this translates into more frequent close fly-bys with the 
     innermost planet around that time (see Fig. 
     \ref{distancesVA}). 
     \hfil\par
%
%
%
%
%
%
%
%
     Apart from Mercury that has no known co-orbital companions, Venus still remains as the inner planet that hosts the smallest known 
     number of co-orbitals. From an observational point of view, the scarcity of detected co-orbitals for the two innermost planets is not 
     surprising and it does not necessarily translate into a real, physical lack of objects. If the majority of the actual innermost 
     co-orbitals have $H>$ 21 as equation (\ref{law}) suggests and spend most of the time far from the Earth and away from the ecliptic plane 
     following orbits similar to those of the three known Venus co-orbitals, their apparent magnitude may be too faint to be detected by the 
     average telescope used to discover NEOs even at the most favourable geometry; the maximum elongation when the object's angular 
     separation from the Sun is the largest and it is best observed from Earth. It is not by chance that 2012 XE$_{133}$ was discovered in 
     2012 December as it was its closest approach to the Earth since 1940 December when it passed 0.05 au from our planet. It had a close 
     encounter with the Earth on 2012 December 19 at 0.055 au and it passed 0.048 au from Mercury on 2013 March 23. On the other hand, 
     if most Venus co-orbitals are as faint as 2012 XE$_{133}$ but they do not move in paths that get close to Earth's (for example, 
     low eccentricity co-orbitals), their detectability by ground-based surveys is severely compromised. Only space-borne programmes may be 
     successful in this case. 
     \hfil\par
     Asteroid 2012 XE$_{133}$ is part of the list compiled by the JPL Sentry System\footnote{http://neo.jpl.nasa.gov/risk/} that includes 
     objects that may be involved in future Earth impact events. Our calculations indicate that an actual collision with the Earth during 
     the next 10\,000 yr cannot be completely ruled out. On 2028 December 30, 2012 XE$_{133}$ may pass close to the Earth and the Moon at a 
     distance $<$ 0.006 au from both bodies. A similar encounter will take place in 2049; even closer will be the encounter in 2098 at 0.003 
     au. In 2151 April, it will approach 0.0027 au from Earth. Some of our control orbits record close approaches below 0.00068 au. 

  \section{Conclusions}
     In this paper we have identified a new high eccentricity Venus co-orbital, 2012 XE$_{133}$. This identification is consistent with 
     Morais \& Morbidelli (2006) predictions in terms of absolute magnitude. Our analysis of the orbit of 2012 XE$_{133}$ reveals that this 
     small NEO is following a transitional, highly chaotic path. The object has been a co-orbital companion to Venus for several thousand 
     years and it may become a passing object within the next few hundred years. The future orbital evolution of this recently discovered
     minor body is not easily predictable beyond that time-scale. Its orbit is the most chaotic among Venus co-orbitals. The three known Venus 
     co-orbitals follow similar orbits with e-folding times of the order of 100 yr and exhibit resonant (or near-resonant) behaviour with 
     Mercury, Venus and the Earth. Our calculations indicate that the thickness of Venus' co-orbital region is 0.004 au with 2007 AG,
     2002 LT$_ {24}$ and 2003 KO$_2$ signalling the edge of the region. Extrapolation of the number distribution of Venus co-orbitals as a 
     function of the absolute magnitude suggests that dozens of objects similar to 2012 XE$_{133}$ could be transient companions to Venus 
     but ground-based surveys may have substantial difficulties in finding them due to the nature of their orbits and intrinsic faintness. 
     \hfil\par
     Our calculations did not include the Yarkovsky effect which may have a non-negligible role on the medium, long-term evolution of 
     objects as small as 2012 XE$_{133}$. Proper modeling of the Yarkovsky force requires knowledge on the physical properties of the 
     objects involved (for example, rotation rate, albedo, bulk density, surface conductivity, emissivity) which is not the case for 2012 
     XE$_{133}$. Detailed observations during future encounters with the Earth should be able to provide that information. The non-inclusion 
     of this effect has no impact on the assessment of the current status of this object as a Venus co-orbital.

  \section*{Acknowledgments}
     The authors would like to thank S. J. Aarseth for providing the code used in this research and the referee for his/her constructive and 
     useful reports. This work was partially supported by the Spanish 'Comunidad de Madrid' under grant CAM S2009/ESP-1496 (Din\'amica 
     Estelar y Sistemas Planetarios). The authors also thank M. J. Fern\'andez-Figueroa, M. Rego Fern\'andez and the Department of Astrophysics of the 
     Universidad Complutense de Madrid (UCM) for providing excellent computing facilities. Most of the calculations and part of the data 
     analysis were completed on the 'Servidor Central de C\'alculo' of the UCM and the authors thank S. Cano Als\'ua for his help during this stage. 
     In preparation of this paper, the authors made use of the NASA Astrophysics Data System, the ASTRO-PH e-print server and the MPC data server.

\end{document}